\documentclass[preprint,onecolumn,superscriptaddress]{revtex4}
\usepackage{graphicx}
\usepackage{amsmath}
\usepackage{amssymb}

\begin{document}

\title{Long-range superharmonic Josephson current and spin-triplet pairing correlations in a junction with ferromagnetic bilayers}
\author{Hao Meng}
\affiliation{Department of Physics, South University of Science and Technology of China, Shenzhen, 518055, China}
\affiliation{School of Physics and Telecommunication Engineering, Shaanxi University of Technology, Hanzhong 723001, China}
\affiliation{National Laboratory of Solid State Microstructures and Department of Physics, Nanjing University, Nanjing 210093, China}
\author{Jiansheng Wu}
\email{wu.js@sustc.edu.cn}
\affiliation{Department of Physics, South University of Science and Technology of China, Shenzhen, 518055, China}
\author{Xiuqiang Wu}
\affiliation{National Laboratory of Solid State Microstructures and Department of Physics, Nanjing University, Nanjing 210093, China}
\author{Mengyuan Ren}
\affiliation{School of material science and technology, Harbin university of science and technology, Harbin 150080, china}
\author{Yajie Ren}
\affiliation{School of Physics and Telecommunication Engineering, Shaanxi University of Technology, Hanzhong 723001, China}

 \begin{abstract}
 The long-range spin-triplet supercurrent transport is an interesting phenomenon in the superconductor/ferromagnet ($S/F$) heterostructure containing noncollinear magnetic domains. Here we study the long-range superharmonic Josephson current in asymmetric $S/F_1/F_2/S$ junctions. It is demonstrated that this current is induced by spin-triplet pairs $\mid\uparrow\uparrow\rangle-$$\mid\downarrow\downarrow\rangle$ or $\mid\uparrow\uparrow\rangle+$$\mid\downarrow\downarrow\rangle$ in the thick $F_1$ layer. The magnetic rotation of the particularly thin $F_2$ layer will not only modulate the amplitude of the superharmonic current but also realise the conversion between $\mid\uparrow\uparrow\rangle-$$\mid\downarrow\downarrow\rangle$ and $\mid\uparrow\uparrow\rangle+$$\mid\downarrow\downarrow\rangle$. Moreover, the critical current shows an oscillatory dependence on thickness and exchange field in the $F_2$ layer. These effect can be used for engineering cryoelectronic devices manipulating the superharmonic current. In contrast, the critical current declines monotonically with increasing exchange field of the $F_1$ layer, and if the $F_1$ layer is converted into half-metal, the long-range supercurrent is prohibited but $\mid\uparrow\uparrow\rangle$ still exists within the entire $F_1$ region. This phenomenon contradicts the conventional wisdom and indicates the occurrence of spin and charge separation in present junction, which could lead to useful spintronics devices.
\end{abstract}

 \maketitle
  Superconductor/ferromagnet ($S/F$) hybrid structure has recently attracted considerable attention because of the potential applications in spintronics and quantum information~\cite{Buz,BerRMP,Esc} as well as the display of a variety of unusual physical phenomena~\cite{ZoharNussinov,IverBSperstad,MadalinaColci,KueiSun}. In general, if a weak $F$ is adjacent to an s-wave $S$ and there is no interfacial spin-flip scattering, the normal Andreev reflection will generate at $S/F$ interfaces. The process involves an electron incident on the $S/F$ interface from the $F$ at energies less than the superconducting energy gap. The incident electron forms a Cooper pair in the $S$ with the retroreflection of a hole of opposite spin to the incident electron. Consequently, the conventional spin-singlet Cooper pair decays at a short range in ferromagnetic region. In $S/F/S$ Josephson junctions with homogeneous magnetization, through the normal Andreev reflection occurring at two $S/F$ interfaces, a Cooper pair is transferred from one $S$ to another, creating a supercurrent flow across the junction~\cite{AAGolubov}. As a consequence of the exchange splitting of the Fermi level of the $F$, the Cooper pair decay in an oscillatory manner superimposed on an exponential decay in the $F$. Correspondingly, the Josephson current displays a damped oscillation with increasing the thickness or the exchange field of the $F$, leading to the appearance of the so-called ``0-$\pi$ transition''~\cite{Buz,BerRMP}. In general, the normal Andreev reflection will be suppressed by the exchange field of the $F$, so the Josephson current just can transport a short distance.

  In contrast, if one insert a thin spin-active $F$ layer with noncollinear magnetization into the $S/F$ interface, it is found that the noncollinear magnetization can lead to a spin-flip scattering, then the reflected hole has the same spin as the incident electron, which is identified as anomalous Andreev reflection. When this reflection takes place at two $S/F$ interfaces, the parallel spin-triplet Cooper pairs $\mid\uparrow\uparrow\rangle$ are generated in the central $F$ layer and can penetrate into $F$ layer over a long distance unsuppressed by the exchange interaction, so that the proximity effect is enhanced. The induced long-range current manifests itself as a large first harmmonic ($I_1\gg{I_2}$) in the spectral decomposition of the Josephson current-phase relation $I(\phi)=I_1\sin(\phi)+I_2\sin(2\phi)+\cdots$~\cite{AAGolubov}.

  It is worth to point that, if the central $F$ layer is converted into fully spin-polarized half-metal, in which electronic bands exhibit insulating behavior for one spin direction and metallic behavior for the other, the normal Andreev reflection will be inhibited completely due to inability to form a pair in the $S$ and impossibility of single-particle transmission. However, the strength of the anomalous Andreev reflection can not be strongly influenced by the spin-polarization of the $F$, and the transport processes of $\mid\uparrow\uparrow\rangle$ (or $\mid\downarrow\downarrow\rangle$) in the $F$ region will continue to take place. In response, several different inhomogeneous configurations have been proposed for studying such enhanced proximity effect~\cite{Eschrig,BerPRL,AFVolkov,Asano,Volkov,Mohammad,Halasz}. The corresponding experiments have proved these physical process and observed the strong enhancement of the long-range spin-triplet supercurrents~\cite{Keizer,Anwar,Robinson,Khaire,Sprungmann,Klose}.

  Different from the configurations mentioned above, it has proposed a long-range proximity effect develops in highly asymmetric $S/F_1/F_2/S$ junction composed of thick $F_1$ layer and particularly thin $F_2$ layer with noncollinear magnetizations at low temperatures~\cite{Tri,LukaTrifunovic,Richard}. This effect arises from two normal Andreev reflections occurred at normal $S/F_1$ interface and two anomalous Andreev reflections at spin-active $F_2/S$ interface. The long-range spin-triplet correlations in this junction give the dominant second harmonic ($I_2\gg{I_1}$) in current-phase relation~\cite{LukaTrifunovic}, which is known as superharmonic Josephson current~\cite{Tri}. Recently, Iovan \emph{et al.}~\cite{AdrianIovan} experimentally observed the long-range supercurrent through above junction. This second harmonic can be manifested as half-integer Shapiro steps that can be experimentally observed~\cite{Sellier}, and the two times smaller flux quantum will be obtained, leading to more sensitive quantum interferometers (SQUIDs)~\cite{Radovic}. It should be stressed that Refs.~\cite{Tri,LukaTrifunovic,Richard} did not discuss the difference of long-range triplet pairing fashion between asymmetric $S/F_1/F_2/S$ junction and symmetric $S/F_2/F_1/F_2/S$. Moreover, it is high desirable to clarify the effect of the misorientation angle on the triplet pairing correlations in the $S/F_1/F_2/S$ junction, as well as the influence of the thickness and the exchange field in two ferromagnetic layers on the Josephson current and the long-range spin-triplet correlations.

  In this work, we study the relation between the long-range superharmonic Josephson current and the spin-triplet pairing correlations in $S/F_1/F_2/S$ junction. It is proposed that the superharmonic Josephson current is induced by the spin-triplet pairs $\mid\uparrow\uparrow\rangle-$$\mid\downarrow\downarrow\rangle$ or $\mid\uparrow\uparrow\rangle+$$\mid\downarrow\downarrow\rangle$ in the long $F_1$ layer. The variation of the misorientation angle between two magnetizations will not only turn the amplitude of the superharmonic current but also realize the conversion between $\mid\uparrow\uparrow\rangle-$$\mid\downarrow\downarrow\rangle$ and $\mid\uparrow\uparrow\rangle+$$\mid\downarrow\downarrow\rangle$. This can be used to control the superharmic current and the pairing fashion in the $F_1$ layer through modulating the magnetic structure of the $F_2$ layer. Besides, the critical current shows an oscillatory dependence on the thickness and exchange field of the highly thin $F_2$ layer. These effect can be used for engineering cryoelectronic devices manipulating spin-polarized supercurrent. In contrast, the critical current decreases monotonically with increasing exchange field of the $F_1$ layer. Specifically, if the $F_1$ layer is converted into half-metal, the long-range Josephson current will be completely prohibited, but $\mid\uparrow\uparrow\rangle$ still exist in $F_1$ region. This phenomenon indicates the occurrence of spin and charge separation in present $S/F$ junction which could lead to useful spintronics devices. These results also contradict the traditional view: the long-range Josephson current is determined by the parallel spin-triplet pairs in the multilayer junction with noncollinear magnetization alignment between ferromagnetic layers. At last, it is also found that the magnetization of the $F_2$ layer will bring about a same direction magnetization in the $F_1$ layer on condition that the magnetic moment of the $F_1$ layer is weak.

\begin{figure}[ptb]
  \centering
  \includegraphics[width=3.1in]{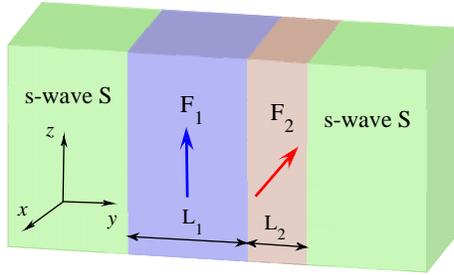} 
  \caption{\textbf{Schematic illustration of the $S/F_1/F_2/S$ Josephson junction containing a bilayer ferromagnet.} Thick arrows in $F_1$ layer and $F_2$ layer indicate the directions of the magnetic moments. The phase difference between the two s-wave $S$s is $\phi=\phi_R-\phi_L$.}
  \label{fig.1}
  \end{figure}

  To be more precise, we consider the Josephson junction consists of two s-wave superconducting electrodes and ferromagnetic bilayer with noncollinear magnetizations. The schematic picture of the $S/F_1/F_2/S$ device is presented in Fig.~\ref{fig.1}. One assume that the transport direction is along the \emph{y} axis, and the system satisfies translational invariance in the \emph{x-z} plane. The thicknesses of $F_1$ layer and $F_2$ layer are $L_{1}$ and $L_{2}$, respectively. The exchange field $\vec{h}$ due to the ferromagnetic magnetizations in the $F_p$ ($p=1, 2$) layer is described by $\vec{h}=h_p(\sin\theta_p\cos\varphi_p, \sin\theta_p\sin\varphi_p, \cos\theta_p)$. Here $\theta_p$ is the tilt angle from the $z$ axis, and $\varphi_p$ is the horizontal angle respect to $x$ axis.

 \section*{Results}

  Based on the extended the Blonder-Tinkham-Klapwijk (BTK) approach~\cite{Blo,Fur,Zhe,Tan}, the dc Josephson current in the $S/F_1/F_2/S$ junction can be expressed as follows
  \begin{equation}\label{Eq1}
  \begin{aligned}
  &I_{e}(\phi)=\frac{k_BTe\Delta}{4\hbar}\sum_{k_{\parallel}}\sum_{\omega_{n}}\frac{k_{e}(\omega_{n})+k_{h}(\omega_{n})}{\Omega_{n}}\\
  &\times[\frac{a_1(\omega_{n},\phi)-a_2(\omega_{n},\phi)}{k_{e}}+\frac{a_3(\omega_{n},\phi)-a_4(\omega_{n},\phi)}{k_{h}}],
  \end{aligned}
  \end{equation}
  where $\omega_{n}=\pi{k_B}T(2n+1)$ are the Matsubara frequencies with $n=0, 1, 2,\ldots$ and $\Omega_{n}=\sqrt{\omega^{2}_{n}+\Delta^{2}(T)}$. $k_{e(h)}(\omega_{n})$ are the perpendicular components of the wave vectors for electron-like (hole-like) quasiparticles in superconducting regions, and $a_j(\omega_{n},\phi)$ with $j=1, 2, 3, 4$ are the scattering coefficients of the normal Andreev reflection under the condition of four different incoming quasiparticles, electron-like quasiparticles (ELQs) and hole-like quasiparticles (HLQs) with spin up and spin down. Then the critical current is derived from $I_c=max_{\phi}|I_e(\phi)|$.

  By applying the Bogoliubov's self-consistent field method~\cite{Gen,Ketterson}, the triplet pair amplitudes are defined as follows~\cite{Halterman}:
  \begin{equation}\label{Eq2}
   f_{0}(y,t)=\frac{1}{2}\sum_{n}\sum_{qq'}(u^{\uparrow}_{nq}v^{\downarrow*}_{nq'}+u^{\downarrow}_{nq}v^{\uparrow*}_{nq'})\zeta_{q}(y)\zeta_{q'}(y)\eta_{n}(t),
  \end{equation}
  \begin{equation}\label{Eq3}
   f_{1}(y,t)= f_{\uparrow\uparrow}(y,t)-f_{\downarrow\downarrow}(y,t),
  \end{equation}
  \begin{equation}\label{Eq4}
   f_{2}(y,t)= f_{\uparrow\uparrow}(y,t)+f_{\downarrow\downarrow}(y,t),
  \end{equation}
  where $\eta_{n}(t)=\cos(E_nt)-i\sin(E_nt)\tanh(E_n/2k_BT)$, and equal-spin pair amplitude will be denoted by $f_{\alpha\alpha}(y,t)=\frac{1}{2}\sum_{n}\sum_{qq'}u^{\alpha}_{nq}v^{\alpha*}_{nq'}\zeta_{q}(y)\zeta_{q'}(y)\eta_{n}(t)$. The singlet pair amplitude writes as $f_3(y)=\Delta(y)/g(y)$. In this paper, the singlet and triplet pair amplitudes are all normalized to the value of the singlet pairing amplitude in a bulk superconducting material. The LDOS is given by~\cite{Halterman}
  \begin{equation}\label{Eq5}
  \begin{aligned}
  N(y,\epsilon)&=-\sum_{n}\sum_{qq'}[(u^{\uparrow}_{nq}u^{\uparrow*}_{nq'}+u^{\downarrow}_{nq}u^{\downarrow*}_{nq'})f'(\epsilon-E_{n})\\
  &+(v^{\uparrow}_{nq}v^{\uparrow*}_{nq'}+v^{\downarrow}_{nq}v^{\downarrow*}_{nq'})f'(\epsilon+E_{n})]\zeta_{q}(y)\zeta_{q'}(y),
  \end{aligned}
  \end{equation}
  where $f'(\epsilon)=\partial{f}/\partial{\epsilon}$ is the derivative of the Fermi function. The LDOS is normalized to unity in the normal state of the $S$ material. In addition, the local magnetic moment in the $S/F_1/F_2/S$ geometry has three components~\cite{Halterman}
  \begin{equation}\label{Eq6}
  \begin{aligned}
   M_{x}(y)&=-\mu_{B}\sum_{n}\sum_{qq'}[(u^{\uparrow*}_{nq}u^{\downarrow}_{nq'}+u^{\downarrow*}_{nq}u^{\uparrow}_{nq'})f_n\\
   &+(v^{\uparrow}_{nq}v^{\downarrow*}_{nq'}+v^{\downarrow}_{nq}v^{\uparrow*}_{nq'})(1-f_n)]\zeta_{q}(y)\zeta_{q'}(y),
  \end{aligned}
  \end{equation}
  \begin{equation}\label{Eq7}
  \begin{aligned}
   M_{y}(y)&=i\mu_{B}\sum_{n}\sum_{qq'}[(u^{\uparrow*}_{nq}u^{\downarrow}_{nq'}-u^{\downarrow*}_{nq}u^{\uparrow}_{nq'})f_n\\
   &+(v^{\uparrow}_{nq}v^{\downarrow*}_{nq'}-v^{\downarrow}_{nq}v^{\uparrow*}_{nq'})(1-f_n)]\zeta_{q}(y)\zeta_{q'}(y),
  \end{aligned}
  \end{equation}
  \begin{equation}\label{Eq8}
  \begin{aligned}
   M_{z}(y)&=-\mu_{B}\sum_{n}\sum_{qq'}[(u^{\uparrow*}_{nq}u^{\uparrow}_{nq'}-u^{\downarrow*}_{nq}u^{\downarrow}_{nq'})f_n\\
   &+(v^{\uparrow}_{nq}v^{\uparrow*}_{nq'}-v^{\downarrow}_{nq}v^{\downarrow*}_{nq'})(1-f_n)]\zeta_{q}(y)\zeta_{q'}(y),
  \end{aligned}
  \end{equation}
  where $\mu_{B}$ and $f_n$ are the Bohr magneton and the Fermi function, respectively. It is convenient to normalize these components to $-\mu_{B}$.

  \begin{figure}[htb]
  \centering
  \includegraphics[width=3.1in]{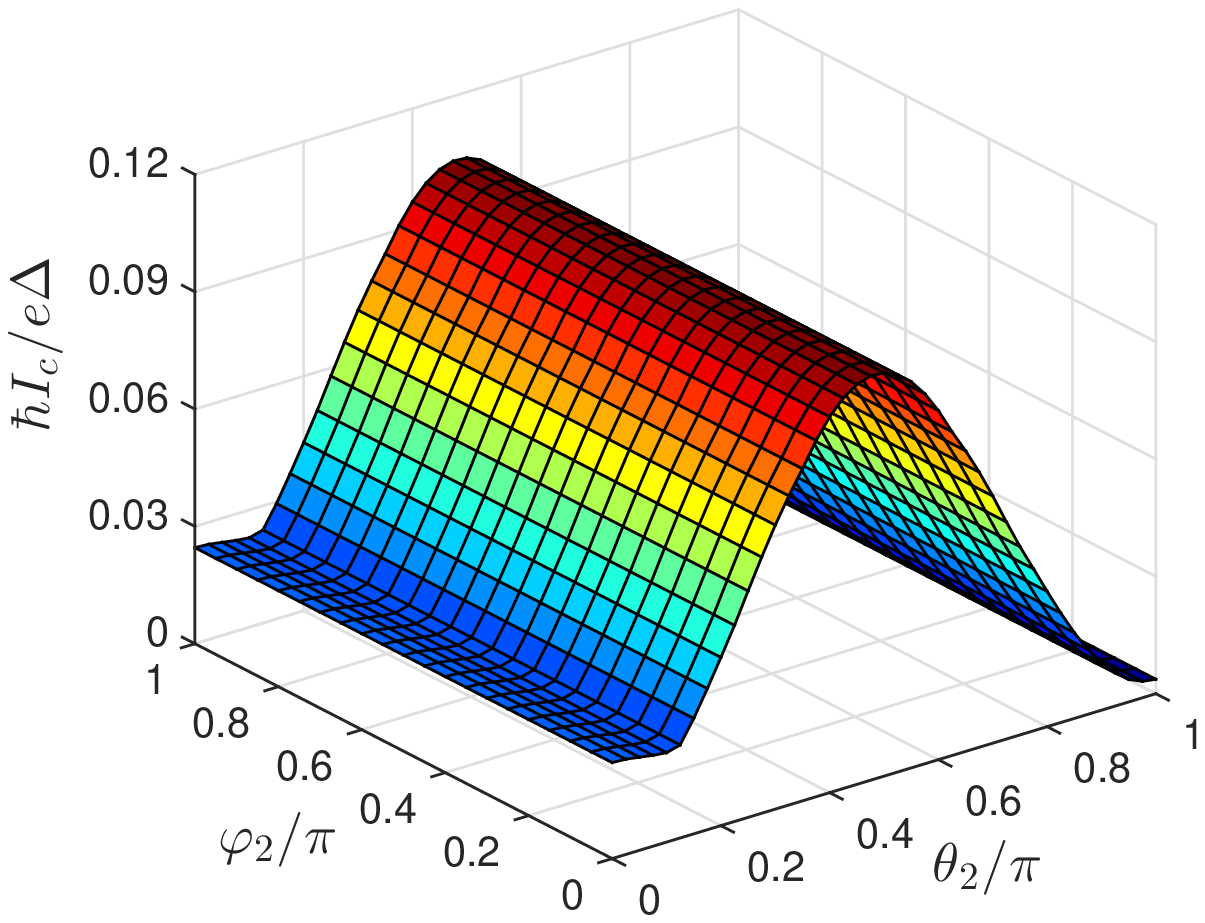} 
  \caption{\textbf{Critical current as a function of the orientation angle ($\theta_2$, $\varphi_2$) of the $F_2$ layer.} Here we set $k_FL_{1}=200$, $k_FL_{2}=6$, $h_1/E_F=0.1$, and $h_2/E_F=0.16$.}
  \label{fig.2}
  \end{figure}

  Unless otherwise stated, in BTK approach we use the superconducting gap $\Delta_0$ as the unit of energy. The Fermi energy is $E_F=1000\Delta_0$, the interface transparency is $Z_{1\text{--}4}=0$ and $T/T_c=0.1$. We measure all lengths and the exchange field strengths in units of the inverse of the Fermi wave vector $k_F$ and the Fermi energy $E_F$, respectively. The magnetization in the $F_1$ layer is fixed along the $z$ direction ($\theta_1=0$, $\varphi_1=0$), while the $F_2$ is a free layer in which the magnetization points any direction. In Bogoliubov's self-consistent field method, we consider the low-temperature limit and take $k_FL_{S1}=k_FL_{S2}=400$, $\omega_D/E_F=0.1$. The other parameters are the same as the ones mentioned before.

   \section*{Discussion}
   \subsection{Superharmonic currents versus misalignment angle}
   \begin{figure}[htb]
   \centering
   \includegraphics[width=3.6in]{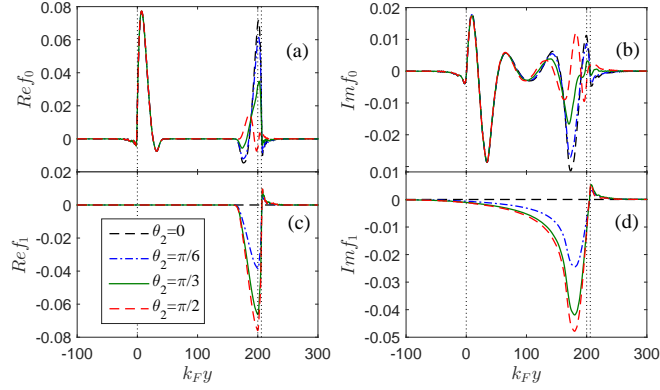} 
   \caption{\textbf{The spin-triplet pair amplitudes $f_0$ and $f_{1}$ plotted as a function of the coordinate $k_Fy$ for several values of $\theta_2$ in the case of $\varphi_2=0$.} The left panels show the real parts while the right ones show the imaginary parts. The dotted vertical lines represent the location of the $S/F_1$, $F_1/F_2$ and $F_2/S$ interfaces. Here $k_FL_{1}=200$, $k_FL_{2}=6$, $h_1/E_F=0.1$, $h_2/E_F=0.16$, $\omega_Dt=4$, and $\phi=0$. All panels utilize the same legend.}
   \label{fig.3}
   \end{figure}
    From Fig.~\ref{fig.2} one can clearly see that the critical current reaches maximum for perpendicular magnetizations ($\theta_2=\pi/2$) and decreases to minimum as the magnetizations are parallel ($\theta_2=0$) or antiparallel ($\theta_2=\pi$) to each other. However, the variation of the angle $\varphi_2$ can not lead to the change of critical current while keeping $\theta_2$ constant.  It is known that characteristic variations of the critical current $I_c$ with the misaligned angles ($\vartheta_2$, $\varphi_2$) are related to the nature of pairing correlations. Figure~\ref{fig.3} shows the spatial distribution of the spin-triplet pair amplitudes for different misalignment angle $\theta_2$ at fixed $\varphi_2=0$. It is found that the real part of $f_0$ and $f_1$ can not penetrate entire $F_1$ layer, but their image parts can be distributed throughout this region. With increasing $\theta_2$, the left parts of $Imf_0$ are almost unchanged, however, their right parts gradually decrease. Correspondingly, the amplitudes of $Imf_1$ increase and turn to maximum at $\theta_2=\pi/2$. The main reason is because the $x$-projection of misaligned magnetic moment in the $F_2$ layer can generate two separate effects: spin-mixing and spin-flip scattering process~\cite{Eschrig}. The former will result a mixture of singlet pairs and triplet pairs with zero spin projection $(\mid\uparrow\downarrow\rangle-$$\mid\downarrow\uparrow\rangle)_x\cos(Q\cdot{R})+i(\mid\uparrow\downarrow\rangle+$$\mid\downarrow\uparrow\rangle)_x\sin(Q\cdot{R})$, where $Q\simeq2h/\hbar{v_F}$, $v_F$ is the Fermi velocity and $R$ is the distance from the $F_2/S$ interface. The latter can convert $(\mid\uparrow\downarrow\rangle+$$\mid\downarrow\uparrow\rangle)_x$ into the parallel spin-triplet pairs $(\mid\uparrow\uparrow\rangle-$$\mid\downarrow\downarrow\rangle)_z$~\cite{Esc}. These parallel spin pairs will penetrate coherently over a long distance into the $F_1$ layer. So the transport of $(\mid\uparrow\uparrow\rangle-$$\mid\downarrow\downarrow\rangle)_z$ can make a significant contribution to superharmonic Josephson current. Meanwhile, the period of this current becomes $\pi$ and satisfies the second harmonic current-phase relation $I_e(\phi)\propto{\sin2\phi}$~\cite{Tri,Richard}. By contrast, in the Josephson junction with ferromagnetic trilayer only spin-triplet pairs $\mid\uparrow\uparrow\rangle$ (or $\mid\downarrow\downarrow\rangle$) can transmit in central ferromagnetic layer, which provide the main contribution to the long-range first harmonic current~\cite{MHouzet}.

   \begin{figure}[htb]
   \centering
   \includegraphics[width=2.6in]{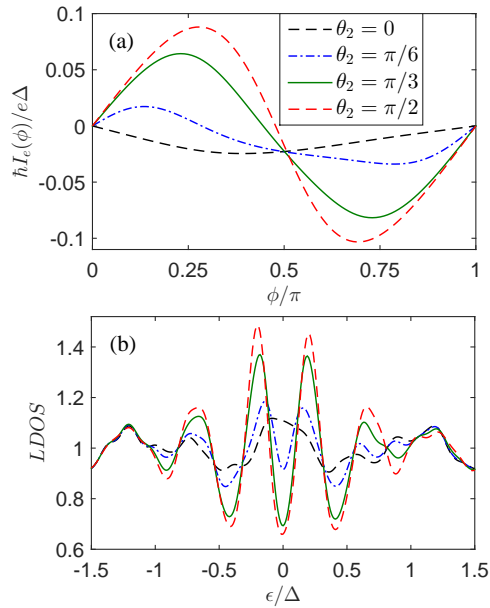} 
   \caption{(a) the Josephson current-phase relation $I_e(\phi)$ for four values of the relative angle $\theta_2$ between magnetizations. (b) The normalized LDOS in the $F_1$ layer ($k_Fy =180$) plotted versus the dimensionless energy $\epsilon/\Delta$ for different $\theta_2$, and the results are calculated at $k_BT=0.0008$. Other parameters are the same as in Fig.~\ref{fig.3}.}
   \label{fig.4}
   \end{figure}

\begin{figure}[htb]
   \centering
   \includegraphics[width=3.6in]{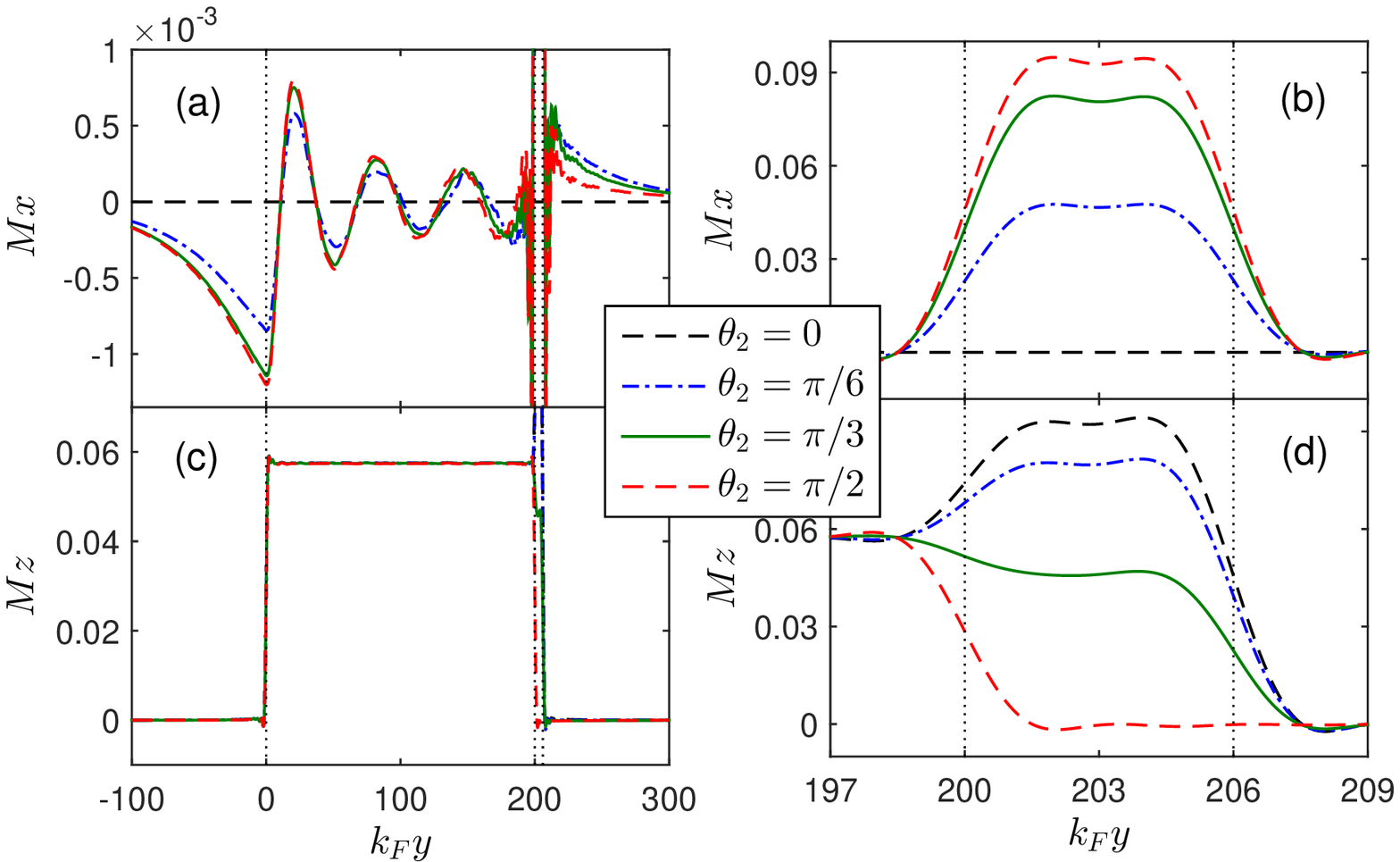} 
   \caption{\textbf{The $x$ (top panels) and $z$ components (bottom panels) of the local magnetic moment plotted as a function of the coordinate $k_Fy$ for different $\theta_2$.} The left panels show the behaviours over the extended $F_1$ regions while the right ones show the detailed behaviours in the $F_2$ layer. Other parameters are the same as in Fig.~\ref{fig.3}.}
   \label{fig.5}
   \end{figure}

   As plotted in Fig.~\ref{fig.4}, in the case of collinear orientation of magnetizations ($\theta_2=0$), the current $I_e(\phi)$ is weak enough and present a first harmonic feature. At this time, the long-range spin-triplet pairs $\mid\uparrow\uparrow\rangle-$$\mid\downarrow\downarrow\rangle$ are absent, so the LDOS in the $F_1$ layer is almost equal to its normal metal value. With increasing $\theta_2$, the magnitude of the second harmonic current is enhanced by the increased number of $\mid\uparrow\uparrow\rangle-$$\mid\downarrow\downarrow\rangle$. Specifically, for orthogonal magnetizations ($\theta_2=\pi/2$), the second harmonic current grows big enough. Correspondingly, the LDOS is significantly enhanced with two distinguishable peaks. Moreover, the spatial profile of the local magnetic moments are plotted for several values of $\theta_2$ in Fig.~\ref{fig.5}. What's most interesting is that the component $M_x$ grows very quickly in the $F_2$ region with increasing $\theta_2$, and also displays the penetration of the same component into the $F_1$ region. The induced $M_x$ in the $F_1$ region tends to not only change magnitude as a function of position, but it also rotates direction. However, the component $M_z$ in the $F_2$ region will gradually decrease with $\theta_2$ and remains almost unchanged in $F_1$ region.

\begin{figure}[htb]
   \centering
   \includegraphics[width=3.6in]{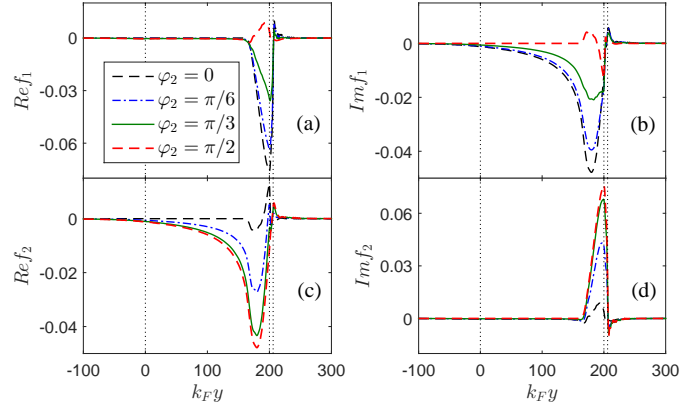} 
   \caption{\textbf{The spin-triplet pair amplitudes $f_{1}$ [(a) and (b)] and $f_{2}$ [(c) and (d)] plotted as a function of the coordinate $k_Fy$ for several values of $\varphi_2$ in the case of $\theta_2=\pi/2$.} The left panels [(a) and (c)] show the real parts while the right ones [(b) and (d)] show the imaginary parts. Other parameters are the same as in Fig.~\ref{fig.3}.}
   \label{fig.6}
   \end{figure}

\begin{figure}[htb]
   \centering
   \includegraphics[width=3.6in]{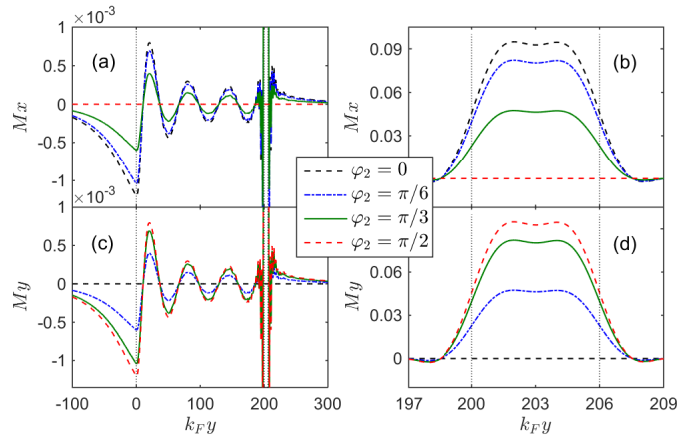} 
   \caption{\textbf{The $x$ (top panels) and $y$ components (bottom panels) of the local magnetic moment plotted as a function of the coordinate $k_Fy$ for different $\varphi_2$.} The left panels show the behaviours over the extended $F_1$ region while the right ones show the detailed behaviours in the $F_2$ region. Other parameters are the same as in Fig.~\ref{fig.3}.}
   \label{fig.7}
   \end{figure}

   As stated above, the variation of the horizontal angle $\varphi_2$ can not influence the Josephson current as the tilt angle $\theta_2$ has a fixed value. However, the change of $\varphi_2$ will induced a conversion of pairing fashion in the $F_1$ region. As shown in Fig.~\ref{fig.6}, on the condition of $\theta_2=\pi/2$, $Imf_1$ decrease gradually from a finite value to zero with increasing $\varphi_2$, but $Ref_2$ exhibit the opposite characteristics. These phenomena can be explained as follows: since the magnetic direction of the $F_2$ layer is oriented along the $x$ axis ($\theta_2=\pi/2$, $\varphi_2=0$), $(\mid\uparrow\downarrow\rangle+$$\mid\downarrow\uparrow\rangle)_x$ in the $F_2$ layer can be converted into $(\mid\uparrow\uparrow\rangle-$$\mid\downarrow\downarrow\rangle)_z$ in the $F_1$ layer. In contrast, if the magnetic moment of the $F_2$ layer is along $y$ axis ($\theta_2=\pi/2$, $\varphi_2=\pi/2$), $(\mid\uparrow\downarrow\rangle+$$\mid\downarrow\uparrow\rangle)_y$ will be transformed into $i(\mid\uparrow\uparrow\rangle+$$\mid\downarrow\downarrow\rangle)_z$, which can also penetrate into the $F_1$ region a long distance and make a major contribution to the second harmonic current. At the same time, when the magnetization direction of the $F_2$ layer rotate from the $x$ axis to the $y$ axis, the induced magnetic moment in the $F_1$ layer would correspondingly turn from $M_x$ to $M_y$, as seen in Fig.~\ref{fig.7}. In what follows, we focus on the dependence of the critical current on the thickness and exchange fields of two ferromagnetic layers under the condition of $\varphi_2=0$.

   \subsection{Superharmonic currents versus thickness and exchange field of the spin-active $F_2$ layer} 
\begin{figure}[htb]
   \centering
   \includegraphics[width=3.2in]{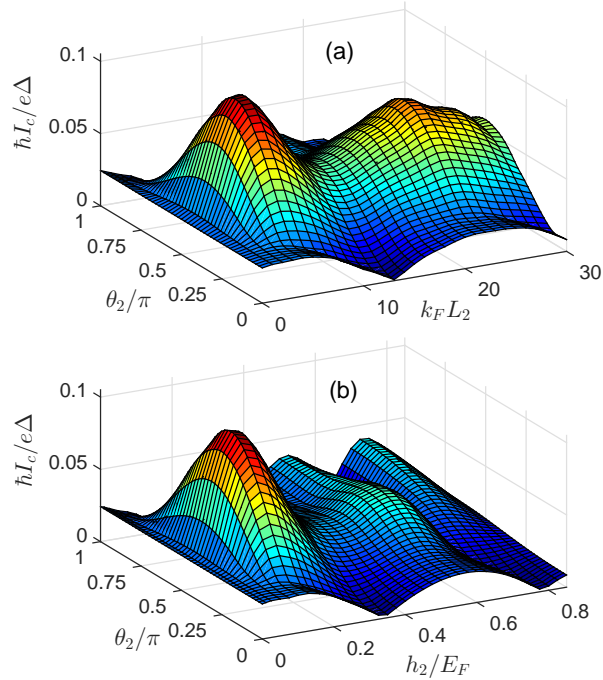} 
   \caption{Critical current (a) as a function of $k_FL_{2}$ and $\theta_2$ for $h_2/E_F=0.16$, and (b) as a function of $h_2/E_F$ and $\theta_2$ for $k_{F}L_{2}=6$. We set $k_FL_1=200$, $h_1/E_F=0.1$, and $\varphi_2=0$.}
   \label{fig.8}
   \end{figure}

Figure~\ref{fig.8} shows the dependence of the critical current $I_c$ on the length $k_FL_{2}$ and exchange field $h_2/E_F$ for different misalignment angle $\theta_2$ when the $F_1$ layer has fixed values $h_1/E_F=0.1$ and $k_FL_{1}=200$. One can see that $I_c$ is sufficiently weak and decays in an oscillatory manner in parallel ($\theta_2=0$) and antiparallel ($\theta_2=\pi$) alignments of the magnetizations. This is because the exchange field in the $F_2$ layer induces a splitting of the energy bands for spin up and spin down. This effect can make $I_c$ oscillate with a period $2\pi\xi_{F}$ and simultaneously decay exponentially on the length scale of $\xi_{F}$~\cite{Buz}. Here, $\xi_{F}$ is the magnetic coherence length. In this case, only the spin-singlet pairs $\mid\uparrow\downarrow\rangle-$$\mid\downarrow\uparrow\rangle$ and spin-triplet pairs $\mid\uparrow\downarrow\rangle+$$\mid\downarrow\uparrow\rangle$ exist in the ferromagnetic layer. These two types of pairs can be suppressed by the exchange field of ferromagnetic layer and mainly provide the contribution to the first harmonic current.

   On the other hand, if the orientations of the magnetic moments are perpendicular to each other ($\theta_2=\pi/2$), $I_c$ also displays the oscillated behaviour with increasing $k_FL_2$, but its order of magnitude is larger than for collinear magnetizations. This characteristic behaviour can be attributed to the spatial oscillations of $\mid\uparrow\downarrow\rangle+$$\mid\downarrow\uparrow\rangle$ in the $F_2$ region with period $Q\cdot{R}$. It is well known that the Cooper pair in the $F_2$ layer will acquire a total momentum $Q$ because of the spin splitting of the energy bands. As described in Ref.~\cite{Meng}, for a fixed $Q$ the amplitude of $\mid\uparrow\downarrow\rangle+$$\mid\downarrow\uparrow\rangle$ will vary with the length $R$ ($=k_FL_2$) of the $F_2$ layer. As a result, the oscillated $\mid\uparrow\downarrow\rangle+$$\mid\downarrow\uparrow\rangle$ can be converted into $\mid\uparrow\uparrow\rangle-$$\mid\downarrow\downarrow\rangle$ in the $F_1$ layer by the spin-flip scattering, and then $\mid\uparrow\uparrow\rangle-$$\mid\downarrow\downarrow\rangle$ can propagate over long distance in the $F_1$ layer and lead to the enhanced superharmonic current. Similarly, if one fixes $k_FL_2$ and changes the $h_2/E_F$, the same features about the critical current can be obtained (see Fig.~\ref{fig.8} (b)). It is worth mentioning that this oscillatory behaviour could be different from the oscillation of the critical current with the thickness of $F_2$ layer in $S/F_2/F_1/F_2/S$ junction~\cite{Meng}, because the supercurrent in the central $F_1$ layer derives from the contribution of $\mid\uparrow\uparrow\rangle$ and manifests itself as a dominant first harmonic in the Josephson current-phase relation.

   \begin{figure}[htb]
   \centering
   \includegraphics[width=3.2in]{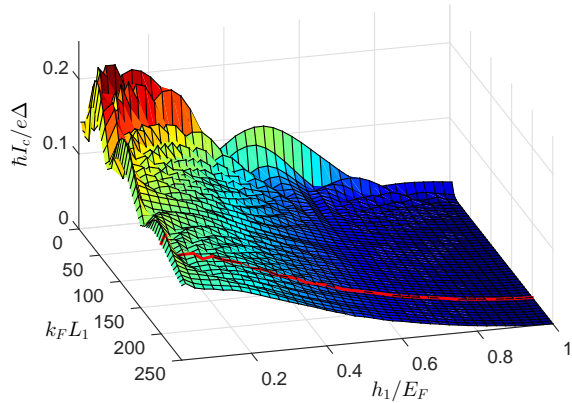} 
   \caption{\textbf{Critical current as a function of $h_1/E_F$ and $k_FL_{1}.$} We set $k_FL_2=6$, $h_2/E_F=0.16$, $\theta_2=\pi/2$, and $\varphi_2=0$.}
   \label{fig.9}
   \end{figure}

   \subsection{Superharmonic currents versus length and exchange field of the long $F_1$ layers} In Fig.~\ref{fig.9} the dependence of the critical current $I_c$ on exchange field $h_1/E_F$ and length $k_FL_1$ are plotted for $\theta_2=\pi/2$. Compared with the Josephson junctions with homogeneous magnetization, $I_c$ in this asymmetric junctions decreases slowly with increasing $k_FL_1$ on the weak or moderate exchange fields. This feature illustrates that $\mid\uparrow\uparrow\rangle-$$\mid\downarrow\downarrow\rangle$ will propagate coherently over long distances in the $F_1$ layer. Furthermore, $I_c$ are almost monotonically decreasing with $h_1/E_F$ for various $k_FL_1$ and will be prohibited completely at $h_1/E_F=1$. It indicates that the superharmonic current will be suppressed by the exchange field of the $F_1$ layer. This phenomenon is clearly different from the first harmonic current in the half-metal Josephson junction with interface spin-flip scattering~\cite{Eschrig,Keizer}, because the first harmonic current induced by $\mid\uparrow\uparrow\rangle$ can not be suppressed by the exchange splitting.

   \begin{figure}[htb]
   \centering
   \includegraphics[width=3.6in]{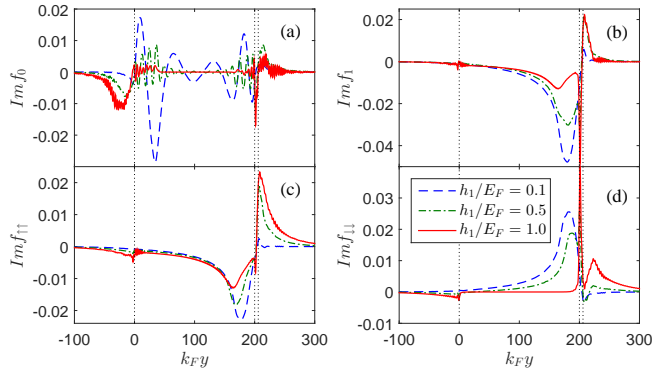} 
   \caption{The imaginary parts of $f_{0}$ (a), $f_{1}$ (b), $f_{\uparrow\uparrow}$ (c) and $f_{\downarrow\downarrow}$ (d) plotted as a function of the coordinate $k_Fy$ for several $h_1/E_F$. We set $k_FL_{1}=200$, $k_FL_{2}=6$, $h_2/E_F=0.16$, $\theta_2=\pi/2$, $\varphi_2=0$, $\omega_Dt=4$, and $\phi=0$.}
   \label{fig.10}
   \end{figure}

\begin{figure}[htb]
   \centering
   \includegraphics[width=2.4in]{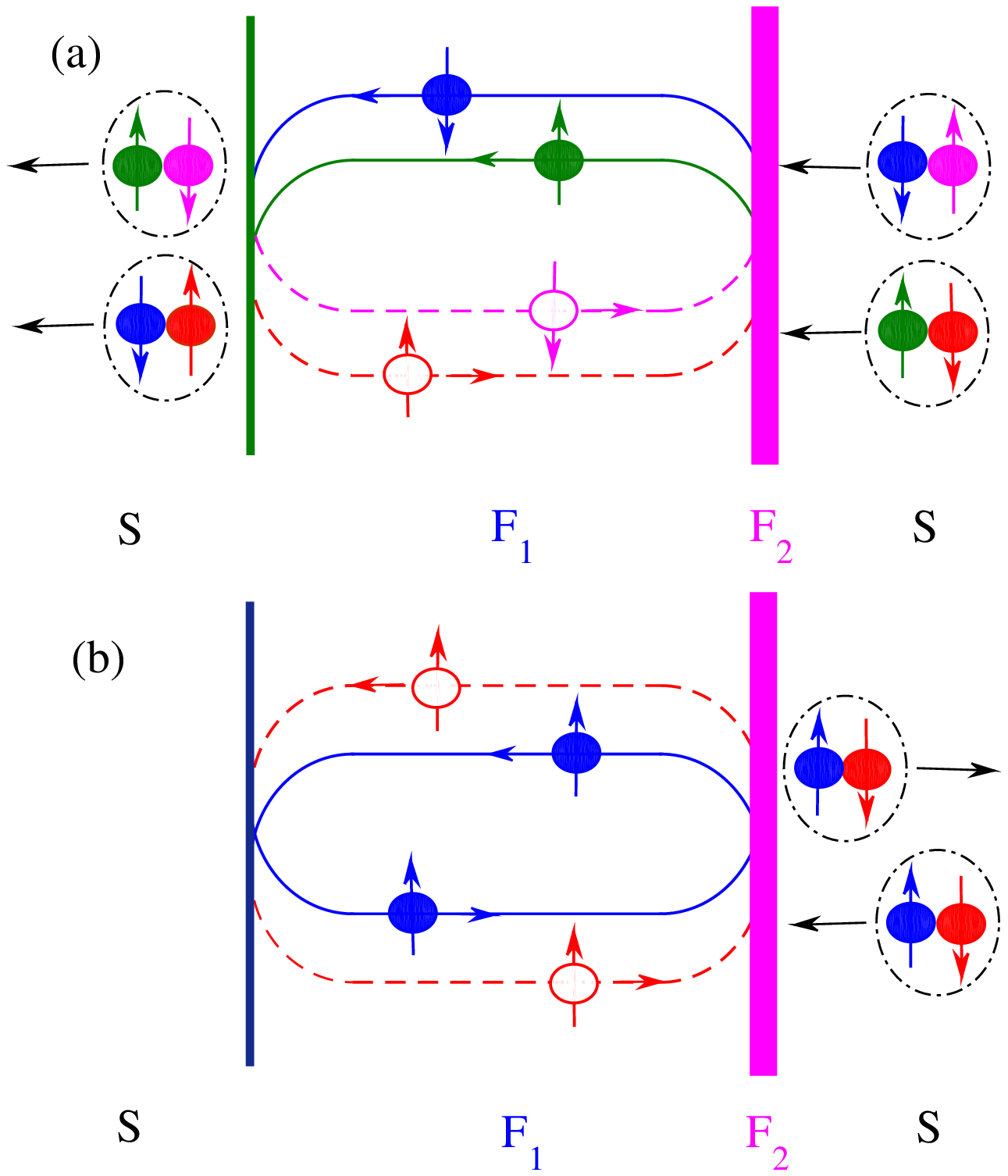} 
   \caption{\textbf{Two types of transference about the pairs of correlated electrons and holes.} (a) The first one consists of two normal Andreev reflections occurred at $S/F_1$ interface and two anomalous Andreev reflections at $F_2/S$ interface in the case of weak exchange field in the $F_1$ layer. (b) The second one consists of two normal reflections at $S/F_1$ interface and two anomalous Andreev reflections at $F_2/S$ interface while the $F_1$ layer is converted into half-metal.}
   \label{fig.11}
   \end{figure}

   In order to clearly explain the contribution of the spin-triplet pairs to the superharmonic current, we choose a fixed length $k_FL_1=200$ for discussion, as illustrated by the red line in Fig.~\ref{fig.9}. Under such conditions, we plot the distribution of the spin-triplet pairing functions $f_0$, $f_1$, $f_{\uparrow\uparrow}$ and $f_{\downarrow\downarrow}$ for three exchange fields $h_1/E_F=0.1$, $0.5$, and $1.0$ in Fig.~\ref{fig.10}. With increasing $h_1/E_F$, the magnitude of $f_0$ and $f_1$ in the $F_1$ region are all reduced and $f_0$ drops to zero at $h_1/E_F=1$. The reason can be summarized as follows: for weak exchange field $h_1/E_F=0.1$ the triplet correlations $f_{\uparrow\uparrow}$ and $f_{\downarrow\downarrow}$ will generate in the $F_2$ region and then combine into $f_{1}$ in the $F_1$ region. $f_{1}$ decay spatially with approaching the $S/F_1$ interface due to the fact that the pairs $\mid\uparrow\uparrow\rangle$ and $\mid\downarrow\downarrow\rangle$ are recombined into the pairs $\mid\uparrow\downarrow\rangle$ and $\mid\downarrow\uparrow\rangle$ by the normal Andreev reflections. For $h_1/E_F=0.5$, $f_{\uparrow\uparrow}$ and $f_{\downarrow\downarrow}$ near the $F_2/S$ interface are both restrained. By contrast, $f_{\uparrow\uparrow}$ adjacent to the $S/F_1$ interface increases instead. Moreover, because $f_{\downarrow\downarrow}$ on the left side of $F_1$ layer is suppressed, the recombination effect at the $S/F_1$ interface becomes weakened, in which case the superharmonic current will decrease. For a fully spin-polarized half-metal ($h_1/E_F=1$), Fig.~\ref{fig.10}(d) shows that $f_{\downarrow\downarrow}$ will be completely suppressed, but $f_{\uparrow\uparrow}$ does not vanish and it's magnitude seems to be a slight increase in the vicinity of the $S/F_1$ interface (see Fig.~\ref{fig.10}(c)). These characters can be attributed to the contributions from two important phenomena taking place at the $S/F_1$ interface: normal Andreev reflections and normal reflections, as shown in Fig.~\ref{fig.11} (a) and (b), respectively.

    If the exchange field $h_1/E_F$ is weak enough, the normal Andreev reflections will mainly occur at the $S/F_1$ interface, which provide the main contribution to $I_c$. In this case, the number of the pairs $\mid\uparrow\uparrow\rangle$ approximately equal to $\mid\downarrow\downarrow\rangle$, and then $\mid\uparrow\uparrow\rangle$ and $\mid\downarrow\downarrow\rangle$ can combine into $\mid\uparrow\uparrow\rangle-$$\mid\downarrow\downarrow\rangle$. Subsequently, $\mid\uparrow\uparrow\rangle-$$\mid\downarrow\downarrow\rangle$ can be converted into $\mid\uparrow\downarrow\rangle-$$\mid\downarrow\uparrow\rangle$ in the left $S$. With increasing $h_1/E_F$, the normal Andreev reflections are gradually being replaced by the normal reflections, and the difference in the number of $\mid\uparrow\uparrow\rangle$ and $\mid\downarrow\downarrow\rangle$ will enlarge simultaneously. As a result, the transition from $\mid\uparrow\uparrow\rangle-$$\mid\downarrow\downarrow\rangle$ to $\mid\uparrow\downarrow\rangle-$$\mid\downarrow\uparrow\rangle$ occurred at the $S/F_1$ interface will be weakened. In the fully spin-polarized case ($h_1/E_F=1$) the absence of the spin down electrons makes it impossible to generate the normal Andreev reflections at $S/F_1$ interface, and therefore the Josephson current is completely suppressed but $\mid\uparrow\uparrow\rangle$ still exist. As depicted in Fig.~\ref{fig.11} (b), the electron transfer process is analogous to the unconventional equal-spin Andreev-reflection process reported in Ref.~\cite{Visani}. Look at the whole picture, it is easy to understand the above process: $\mid\uparrow\downarrow\rangle$ injecting from the right $S$ is converted into $\mid\uparrow\uparrow\rangle$ in the $F_1$ layer, and $\mid\uparrow\uparrow\rangle$ will be consequently reflected normally back as $\mid\uparrow\uparrow\rangle$ at the $S/F_1$ interface. Then $\mid\uparrow\uparrow\rangle$ is transformed into $\mid\uparrow\downarrow\rangle$ by the spin-flip scattering of the $F_2$ layer. At last, $\mid\uparrow\downarrow\rangle$ transports to the right $S$. In the whole process, none of Coopers can penetrate into the left $S$, so the Josephson current would be suppressed completely.

   \begin{figure}[htb]
   \centering
   \includegraphics[width=2.4in]{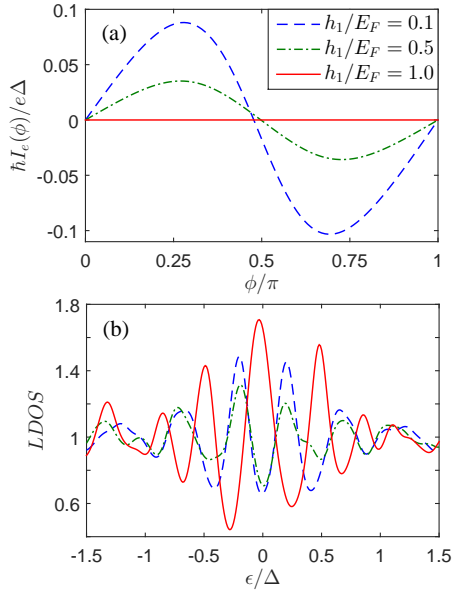} 
   \caption{(a) the Josephson current-phase relation $I_e(\phi)$ for different $h_1/E_F$. (b) The normalized LDOS in the $F_1$ layer ($k_Fy =180$) plotted versus the dimensionless energy $\epsilon/\Delta$, and the results are calculated at $k_BT=0.0008$. Other parameters are the same as in Fig.~\ref{fig.10}.}
   \label{fig.12}
   \end{figure}

\begin{figure}[htb]
   \centering
   \includegraphics[width=3.6in]{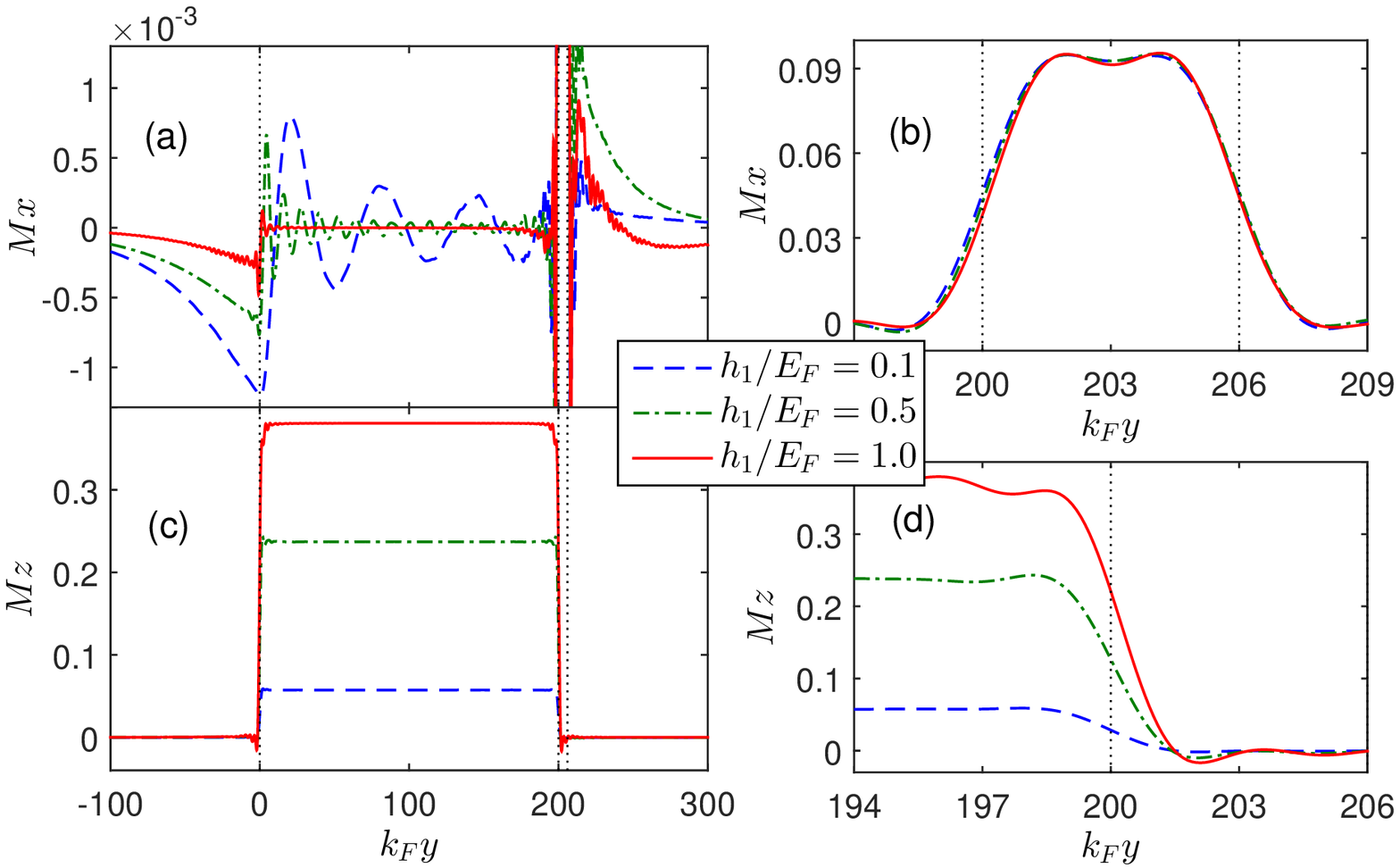} 
   \caption{\textbf{The $x$ (top panels) and $z$ components (bottom panels) of the local magnetic moment plotted as a function  of the coordinate $k_Fy$ for different $h_1/E_F$.} The left panels show the behaviours over the extended $F_1$ region while the right ones show the detailed behaviours in the $F_2$ layer. Other parameters are the same as in Fig.~\ref{fig.10}.}
   \label{fig.13}
   \end{figure}

   In order to facilitate the experimental observations for the future, we plot the current-phase relation and the LDOS in the $F_1$ layer at three points $h_1/E_F=0.1$, $0.5$ and $1.0$ in Fig.~\ref{fig.12}. With increasing $h_1/E_F$, the superharmonic current $I_e(\phi)$ decreases and two distinguishable peaks in the LDOS will become weak correspondingly. It's particularly noteworthy that if $h_1/E_F=1$ Josephson current was completely suppressed but the LDOS displays a sharp zero energy conductance peak which marks the presence of $\mid\uparrow\uparrow\rangle$. It can be measured in principle by STM experiments. And this feature is different from the conventional views: (i) The long-range triplet Josephson current is proportional to the parallel spin-triplet pairs $\mid\uparrow\uparrow\rangle$ or $\mid\downarrow\downarrow\rangle$. (ii) If the long-range triplet supercurrent pass through the Josephson junction, there will present the zero energy conductance peak in the LDOS of $F$. Finally, we discuss the influence of $h_1/E_F$ on the local magnetic moment. As can be seen from Fig.~\ref{fig.13}, in the $F_1$ region $M_z$ will grow with the increase of $h_1/E_F$, but the induced $M_x$ could be suppressed. For $h_1/E_F=1$, the $M_z$ reaches maximum but $M_x$ will disappear. By contrast, $M_x$ in the $F_2$ region hardly changes with $h_1/E_F$, and $M_z$ will partly permeate into the $F_2$ layer.

   To summarize, we have studied the long-range superharmonic Josephson current and the spin-triplet pairing correlations in the asymmetric $S/F_1/F_2/S$ junction. We have shown that the superharmonic current was induced by the spin-triplet pairs $\mid\uparrow\uparrow\rangle-$$\mid\downarrow\downarrow\rangle$ or $\mid\uparrow\uparrow\rangle+$$\mid\downarrow\downarrow\rangle$ in the long $F_1$ layer. The rotation of the magnetic moment in the thin spin-active $F_2$ layer will not only modulate the amplitude of the superharmonic current through the junctions, but also realize the conversion from $\mid\uparrow\uparrow\rangle-$$\mid\downarrow\downarrow\rangle$ to $\mid\uparrow\uparrow\rangle+$$\mid\downarrow\downarrow\rangle$ in the $F_1$ layer. Besides, the critical current oscillates with the length and exchange field in the $F_2$ layer. These features provide an efficient way to control the superharmonic current and the spin-triplet pairing fashion by changing the magnetic moment of the $F_2$ layer. Specifically, the critical current almost decreases monotonically with the exchange field of the $F_1$ layer, and if the $F_1$ layer is converted into half-metal, the Josephson current disappear completely but the spin-triplet pairs $\mid\uparrow\uparrow\rangle$ still exist within the entire $F_1$ layer. This behavior is different from the conventional view about the relationship between the long-range current and the parallel spin-triplet pairs in the junctions with ferromagnetic trilayers. These results therefore indicated that the spin and charge degrees of the freedom can be separated in practice in the junction with ferromagnetic bilayers, and suggested the promising potential of these junctions for spintronics applications.

  \section*{Methods}

  The BCS mean-field effective Hamiltonian is given by~\cite{Buz,Gen}
  \begin{equation}\label{Eq9}
  \begin{aligned}
    H_{eff}&=\int{d\vec{r}}\{\sum_{\alpha,\beta}\psi^{\dag}_{\alpha}(\vec{r})[H_e(\hat{\textbf{1}})_{\alpha\beta}-(\vec{h}\cdot\vec{\sigma})_{\alpha\beta}]\psi_{\beta}(\vec{r})\\
   &+\frac{1}{2}[\sum_{\alpha,\beta}(i\sigma_{y})_{\alpha\beta}\Delta(\vec{r})\psi^{\dag}_{\alpha}(\vec{r})\psi^{\dag}_{\beta}(\vec{r})+h.c.]\},
  \end{aligned}
  \end{equation}
  where $H_e=-\hbar^2\nabla^{2}/2m-E_F$ is the single-particle Hamiltonian, $\psi^{\dag}_{\alpha}(\vec{r})$ and $\psi_{\alpha}(\vec{r})$ are creation and annihilation operators with spin $\alpha$. $\hat{\sigma}$ and $E_F$ denote Pauli matrix and the Fermi energy, respectively. $\Delta(\vec{r})=\Delta(T)[e^{i\phi_{L}}\Theta(-y)+e^{i\phi_{R}}\Theta(y-L_{F})]$ describes the superconducting pair potential with $L_F=L_{1}+L_{2}$. Here $\Delta(T)$ accounts for the temperature-dependent energy gap. It satisfies the BCS relation $\Delta(T)=\Delta_0\tanh(1.74\sqrt{T_c/T-1})$, where $\Delta_0$ is the energy gap at zero temperature and $T_c$ is the superconducting critical temperature. $\Theta(y)$ is the unit step function and $\phi_{L(R)}$ is the phase of the left (right) $S$.

  By making use of the Bogoliubov transformation $\psi_{\alpha}(y)=\sum_{n}[u_{n\alpha}(y)\hat{\gamma}_{n}+v^{\ast}_{n\alpha}(y)\hat{\gamma}^{\dag}_{n}]$ and the anticommutation relations of the quasiparticle annihilation and creation operators $\hat{\gamma}_{n}$ and $\hat{\gamma}^{\dag}_{n}$, we have the Bogoliubov-de Gennes (BdG) equation~\cite{Buz,Gen}
  \begin{equation}\label{Eq10}
   \begin{pmatrix}
   H_e-h_z & -h_x+ih_y & 0 & \Delta(y)      \\
   -h_x-ih_y & H_e+h_z & -\Delta(y) & 0     \\
   0 & -\Delta^{*}(y) & -H_e+h_z & h_x+ih_y \\
   \Delta^{*}(y) & 0 & h_x-ih_y & -H_e-h_z  \\
   \end{pmatrix}
   \begin{pmatrix}
   u_{\uparrow}(y)      \\
   u_{\downarrow}(y)    \\
   v_{\uparrow}(y)      \\
   v_{\downarrow}(y)    \\
  \end{pmatrix}
  =\begin{pmatrix}
   u_{\uparrow}(y)      \\
   u_{\downarrow}(y)    \\
   v_{\uparrow}(y)      \\
   v_{\downarrow}(y)    \\
   \end{pmatrix}.
  \end{equation}

   \textbf{Blonder-Tinkham-Klapwijk approach}
   The BdG equation~(\ref{Eq10}) can be solved for each superconducting electrode and each $F$ layer, respectively. For an incident spin up electron in the left $S$, the wave functions in the $S$ leads and the $F_p$ layer are
  \begin{equation}\label{Eq11}
  \begin{aligned}
  &\Psi^{S}_{L}(y)=(u\hat{e}_{1}e^{\frac{i\phi_{L}}{2}}+v\hat{e}_{4}e^{-\frac{i\phi_{L}}{2}})e^{ik_{e}y} \\
  &+[(a_1\hat{e}_{1}-a'_1\hat{e}_{2})ve^{\frac{i\phi_{L}}{2}}+(a_1\hat{e}_{4}+a'_1\hat{e}_{3})ue^{-\frac{i\phi_{L}}{2}}]e^{ik_{h}y} \\
  &+[(b_1\hat{e}_{1}+b'_1\hat{e}_{2})ue^{\frac{i\phi_{L}}{2}}+(b_1\hat{e}_{4}-b'_1\hat{e}_{3})ve^{-\frac{i\phi_{L}}{2}})e^{-ik_{e}y},
  \end{aligned}
  \end{equation}
  \begin{equation}\label{Eq12}
  \begin{aligned}
  \Psi^{F}_{p}(y)&=T_p\{[f_{p1}{e^{ik^{e\uparrow}_{Fp}y}}+f_{p2}{e^{-ik^{e\uparrow}_{Fp}y}}]\hat{e}_{1}+[f_{p3}{e^{ik^{e\downarrow}_{Fp}y}}\\
  &+f_{p4}{e^{-ik^{e\downarrow}_{Fp}y}}]\hat{e}_{2}+[f_{p5}{e^{-ik^{h\uparrow}_{Fp}y}}+f_{p6}{e^{ik^{h\uparrow}_{Fp}y}}]\hat{e}_{3}\\
  &+[f_{p7}{e^{-ik^{h\downarrow}_{Fp}y}}+f_{p8}{e^{ik^{h\downarrow}_{Fp}y}}]\hat{e}_{4}\},
  \end{aligned}
  \end{equation}
  \begin{equation}\label{Eq13}
  \begin{aligned}
  &\Psi^{S}_{R}(y)=[(c_1\hat{e}_{1}+c'_1\hat{e}_{2})ue^{\frac{i\phi_{R}}{2}}+(c_1\hat{e}_{4}-c'_1\hat{e}_{3})ve^{-\frac{i\phi_{R}}{2}}]e^{ik_{e}y}\\
  &+[(d_1\hat{e}_{1}-d'_1\hat{e}_{2})ve^{\frac{i\phi_{R}}{2}}+(d_1\hat{e}_{4}+d'_1\hat{e}_{3})ue^{-\frac{i\phi_{R}}{2}}]e^{-ik_{h}y}.
  \end{aligned}
  \end{equation}
  Here $\hat{e}_{1}=[1,0,0,0]^{T}$, $\hat{e}_{2}=[0,1,0,0]^{T}$, $\hat{e}_{3}=[0,0,1,0]^{T}$, $\hat{e}_{4}=[0,0,0,1]^{T}$ are basis wave functions. Quasiparticle amplitudes are defined as $u=\sqrt{(1+\Omega/E)/2}$ and $v=\sqrt{(1-\Omega/E)/2}$ with $\Omega=\sqrt{E^2-\Delta^2}$. The perpendicular components of the ELQs (HLQs) wave vector in $S$ leads and $F_p$ layer are given by $k_{e(h)}=\sqrt{2m[E_F+(-)\Omega]/\hbar^2-k^{2}_{\parallel}}$ and  $k^{e(h)\alpha}_{Fp}=\sqrt{2m[E_F+(-)E+\rho_{\alpha}h_{p}]/\hbar^2-k^{2}_{\parallel}}$ with $\rho_{\uparrow(\downarrow)}=1(-1)$, respectively. It is worthy to note that the parallel component $k_{\parallel}$ is conserved in transport processes of the quasiparticles. The matrix can be defined as~\cite{Jin}
  \begin{equation}\label{Eq14}
  \setlength{\arraycolsep}{1.2pt}
  T_p=
  \begin{pmatrix}
   \cos\frac{\theta_p}{2}e^{-i\varphi_p} & -\sin\frac{\theta_p}{2}e^{-i\varphi_p} & 0 & 0 \\
   \sin\frac{\theta_p}{2} & \cos\frac{\theta_p}{2} & 0 & 0    \\
   0 & 0 & \cos\frac{\theta_p}{2}e^{i\varphi_p} & -\sin\frac{\theta_p}{2}e^{i\varphi_p} \\
   0 & 0 & \sin\frac{\theta_p}{2} & \cos\frac{\theta_p}{2} \\
  \end{pmatrix}.
  \end{equation}
  The coefficients $b_{1}$, $b'_{1}$, $a'_{1}$ and $a_{1}$ describe normal reflection, the normal reflection with spin-flip, anomalous Andreev reflection, and normal Andreev reflection, respectively. $f_{pr}$ ($r=1\text{--}8$) are quasiparticles wave function amplitudes in the $F_p$ layer. Likewise, $c_1$, $d_1$, $c'_1$ and $d'_1$ are the quasiparticles transmission amplitudes in the right superconducting electrode. All scattering coefficients can be determined by solving the continuity conditions of the wave function and its derivative at the interface
  \begin{equation}\label{Eq15}
  \begin{aligned}
  &\Psi^{S}_{L}(y_1)=\Psi^{F}_1(y_1),\partial_{y}[\psi^{F}_{1}-\psi^{S}_{L}]|_{y_1}=2k_FZ_1\psi^{F}_{1}(y_1);\\
  &\Psi^{F}_{1}(y_2)=\Psi^{F}_2(y_2),\partial_{y}[\psi^{F}_{2}-\psi^{F}_{1}]|_{y_2}=2k_FZ_2\psi^{F}_{2}(y_2);\\
  &\Psi^{F}_{2}(y_3)=\Psi^{S}_R(y_3),\partial_{y}[\psi^{S}_{R}-\psi^{F}_{2}]|_{y_3}=2k_FZ_3\psi^{S}_{R}(y_3).
  \end{aligned}
  \end{equation}
  Here $Z_1\text{--}Z_3$ are dimensionless parameters describing the magnitude of the interfacial resistances. $y_{1\text{--}3}=0, L_{1}, L_{F}$ are local coordinate values at the interfaces, and $k_F=\sqrt{2mE_F}$ is the Fermi wave vector. From the boundary conditions, we obtain a system of linear equations that yield the scattering coefficients.

  \textbf{Bogoliubov's self-consistent field method}
   We put the $S/F_1/F_2/S$ junction in a one-dimensional square potential well with infinitely high walls, then the eigenvalues and eigenvectors of the BdG equation~(\ref{Eq10}) have the following changes: $E\rightarrow{E_{n}}$ and $[u_{\uparrow}(y),u_{\downarrow}(y),v_{\uparrow}(y),v_{\downarrow}(y)]^{T}\rightarrow[u_{n\uparrow}(y),u_{n\downarrow}(y),v_{n\uparrow}(y),v_{n\downarrow}(y)]^{T}$. Accordingly, the corresponding quasiparticle amplitudes can be expanded in terms of a set of basis vectors of the stationary states~\cite{Landau}, $u_{n\alpha}(y)$$=$$\sum_{q}u^{\alpha}_{nq}\zeta_{q}(y)$ and $v_{n\alpha}(y)=\sum_{q}v^{\alpha}_{nq}\zeta_{q}(y)$ with $\zeta_{q}(y)=\sqrt{2/L}\sin(q{\pi}y/L)$. Here, $q$ is a positive integer and $L=L_{S1}+L_{F}+L_{S2}$. $L_{S1}$ and $L_{S2}$ are the thicknesses of the left and right superconducting electrodes, respectively. The superconducting pair potential in the BdG equation~(\ref{Eq10}) is determined by the self-consistency condition~\cite{Gen}
   \begin{equation}\label{Eq16}
   \Delta(y)=\frac{g(y)}{2}\sum_{n}{'}\sum_{qq'}(u^{\uparrow}_{nq}v^{\downarrow*}_{nq'}-u^{\downarrow}_{nq}v^{\uparrow*}_{nq'})\zeta_{q}(y)\zeta_{q'}(y)\tanh(\frac{E_n}{2k_BT}),
   \end{equation}
   where the primed sum of $E_n$ is over eigenstates corresponding to positive energies smaller than or equal to the Debye cutoff energy $\omega_D$, and the superconducting coupling parameter $g(y)$ is a constant in the superconducting regions and zero elsewhere. Iterations are performed until self-consistency is reached, starting from the stepwise approximation for the pair potential.

  \section*{Acknowledgments}

  This work is supported by the State Key Program for Basic Research of China under Grants No.2011CB922103 and No.2010CB923400, the National Natural Science Foundation of China under Grants No.11174125, No.11074109, No.51106093 and No.11447112, the Scientific Research Program Funded by Shaanxi Provincial Education Department Grant No.12JK0972 and No.15JK1132, the Scientific Research Foundation of Shaanxi University of Technology Grant No.SLG-KYQD2-01.

  \section*{Author contributions}

   H.M. and J.S.W. conceived the research and performed the calculations. All authors contributed to discussion and reviewed the manuscript.

  \section*{Additional information}

  \textbf{Competing financial interests:} The authors declare no competing financial interests.

  \section*{Figure legends}

  Figure 1: \textbf{Schematic illustration of the $S/F_1/F_2/S$ Josephson junction containing a bilayer ferromagnet.} Thick arrows in $F_1$ layer and $F_2$ layer indicate the directions of the magnetic moments. The phase difference between the two s-wave $S$s is $\phi=\phi_R-\phi_L$.

  Figure 2: \textbf{Critical current as a function of the orientation angle ($\theta_2$, $\varphi_2$) of the $F_2$ layer.} Here we set $k_FL_{1}=200$, $k_FL_{2}=6$, $h_1/E_F=0.1$, and $h_2/E_F=0.16$.

  Figure 3: \textbf{The spin-triplet pair amplitudes $f_0$ and $f_{1}$ plotted as a function of the coordinate $k_Fy$ for several values of $\theta_2$ in the case of $\varphi_2=0$.} The left panels show the real parts while the right ones show the imaginary parts. The dotted vertical lines represent the location of the $S/F_1$, $F_1/F_2$ and $F_2/S$ interfaces. Here $k_FL_{1}=200$, $k_FL_{2}=6$, $h_1/E_F=0.1$, $h_2/E_F=0.16$, $\omega_Dt=4$, and $\phi=0$. All panels utilize the same legend.

  Figure 4: (a) the Josephson current-phase relation $I_e(\phi)$ for four values of the relative angle $\theta_2$ between magnetizations. (b) The normalized LDOS in the $F_1$ layer ($k_Fy =180$) plotted versus the dimensionless energy $\epsilon/\Delta$ for different $\theta_2$, and the results are calculated at $k_BT=0.0008$. Other parameters are the same as in Fig.~\ref{fig.3}.

  Figure 5: \textbf{The $x$ (top panels) and $z$ components (bottom panels) of the local magnetic moment plotted as a function of the coordinate $k_Fy$ for different $\theta_2$.} The left panels show the behaviours over the extended $F_1$ regions while the right ones show the detailed behaviours in the $F_2$ layer. Other parameters are the same as in Fig.~\ref{fig.3}.

  Figure 6:  \textbf{The spin-triplet pair amplitudes $f_{1}$ [(a) and (b)] and $f_{2}$ [(c) and (d)] plotted as a function of the coordinate $k_Fy$ for several values of $\varphi_2$ in the case of $\theta_2=\pi/2$.} The left panels [(a) and (c)] show the real parts while the right ones [(b) and (d)] show the imaginary parts. Other parameters are the same as in Fig.~\ref{fig.3}.

  Figure 7: \textbf{The $x$ (top panels) and $y$ components (bottom panels) of the local magnetic moment plotted as a function of the coordinate $k_Fy$ for different $\varphi_2$.} The left panels show the behaviours over the extended $F_1$ region while the right ones show the detailed behaviours in the $F_2$ region. Other parameters are the same as in Fig.~\ref{fig.3}.

  Figure 8: Critical current (a) as a function of $k_FL_{2}$ and $\theta_2$ for $h_2/E_F=0.16$, and (b) as a function of $h_2/E_F$ and $\theta_2$ for $k_{F}L_{2}=6$. We set $k_FL_1=200$, $h_1/E_F=0.1$, and $\varphi_2=0$.

  Figure 9: \textbf{Critical current as a function of $h_1/E_F$ and $k_FL_{1}.$} We set $k_FL_2=6$, $h_2/E_F=0.16$, $\theta_2=\pi/2$, and $\varphi_2=0$.

  Figure 10: The imaginary parts of $f_{0}$ (a), $f_{1}$ (b), $f_{\uparrow\uparrow}$ (c) and $f_{\downarrow\downarrow}$ (d) plotted as a function of the coordinate $k_Fy$ for several $h_1/E_F$. We set $k_FL_{1}=200$, $k_FL_{2}=6$, $h_2/E_F=0.16$, $\theta_2=\pi/2$, $\varphi_2=0$, $\omega_Dt=4$, and $\phi=0$.

  Figure 11: \textbf{Two types of transference about the pairs of correlated electrons and holes.} (a) The first one consists of two normal Andreev reflections occurred at $S/F_1$ interface and two anomalous Andreev reflections at $F_2/S$ interface in the case of weak exchange field in the $F_1$ layer. (b) The second one consists of two normal reflections at $S/F_1$ interface and two anomalous Andreev reflections at $F_2/S$ interface while the $F_1$ layer is converted into half-metal.

  Figure 12: (a) the Josephson current-phase relation $I_e(\phi)$ for different $h_1/E_F$. (b) The normalized LDOS in the $F_1$ layer ($k_Fy =180$) plotted versus the dimensionless energy $\epsilon/\Delta$, and the results are calculated at $k_BT=0.0008$. Other parameters are the same as in Fig.~\ref{fig.10}.

  Figure 13: \textbf{The $x$ (top panels) and $z$ components (bottom panels) of the local magnetic moment plotted as a function  of the coordinate $k_Fy$ for different $h_1/E_F$.} The left panels show the behaviours over the extended $F_1$ region while the right ones show the detailed behaviours in the $F_2$ layer. Other parameters are the same as in Fig.~\ref{fig.10}.

 \end{document}